\documentclass{spie}%
\usepackage{amssymb}
\usepackage{amsfonts}
\usepackage{amsmath}%
\usepackage{graphicx}
\providecommand{\U}[1]{\protect\rule{.1in}{.1in}}

\newtheorem{corollary}[theorem]{Corollary}
\newtheorem{proposition}[theorem]{Proposition}
\begin{document}

\title{The Notion of a Random Gauge and its Interpretation}
\author{John E Gray\\Code Q-31, \\Electromagnetic and Sensor Systems Department, \\Naval Surface Warfare Center Dahlgren, \\18444 FRONTAGE ROAD SUITE 327,\\DAHLGREN VA 22448-5161}
\maketitle

\begin{abstract}
We introduce the concept of a random gauge. We propose two distinct types of
random gauge that can be defined based on the concept of phase noise in
scattering theory. In the context of quantum physics, we discuss a variety of
possible realizations of this concept that can be connected to various
Aharonov-other effects and make some connections with relativity as well.

\end{abstract}

\section{Introduction}

The guiding principle of Einstein's principle of general relativity is that
the laws of physics should be expressed in a form that is independent of the
coordinate frame in which one might choose to express them. Weyl generalized
this concept in the realm of quantum physics by the Dirac operator for \ a
field by introducing a Riemann connection for space-time, so that the partial
derivative is replaced with a covariant derivative which yields a new Dirac
operator\cite{Frankel2004} :%
\begin{equation}
\bar{\partial}\Psi+\frac{1}{4}\Gamma_{ik}^{j}\gamma^{i}\gamma^{k}\Psi.
\end{equation}
Note $\Psi$ is the field, while $\Gamma\Psi$ is the \emph{interaction} term
that tells us how the gravity field interacts with the electron. When one can
neglect the gravitational interaction, \ one can consider a change of field
coordinates $\Psi$ rather than space-time coordinates, e.g. \emph{gauge
transformations.} (For a history, this summary is based upon the overview by
Gross\cite{Gross1992} .)

Quantum mechanics assigns a physical meaning to the absolute value $\left\vert
\Psi^{\alpha}\right\vert $, however it does not assign a meaning to the phase
of $\Psi^{\alpha}$, so $\Psi^{\alpha}=\left\vert \Psi^{\alpha}\right\vert
e^{i\theta_{\alpha}}$, and $\theta_{\alpha}$, the \emph{phase}, has always
been taken to have no physical meaning. Both the Dirac equation and Lagrangian
are invariant under transformations of the form:
\begin{equation}
\Psi\longmapsto e^{i\theta_{\alpha}}\Psi.
\end{equation}
This is a \emph{global} transformation since $\alpha$ is constant. A global
transformation is an invariance of the Dirac equation. Besides a global
transformation, physicists have also considered the effect of local
transformations of $\alpha$, local in this case means that $\alpha=\alpha(x)$
so it depends on the particular space-time coordinates. \ When coordinates are
included in $\alpha$, the Dirac equation and Lagrangian no longer are
invariant under transformations because a $d\alpha$ appears in the
transformation equations. Thus, one could conclude that there is a background
field that interacts with the electron. One can go through the usual argument
that associates a change of frame with a complex line bundle which leads
naturally to the association of the background field with electromagnetism and
yet another Dirac operator:
\begin{equation}
\bar{\partial}_{A}\triangleq\gamma^{i}\left(  \partial_{i}+\omega_{j}\right)
=\bar{\partial}-ie\gamma^{i}A_{j}.
\end{equation}
Quantum mechanics mandates that $A$ be considered as a separate field, so one
gets a new Lagrangian associated with $A$. This brief review of the power of
invariance associated with the gauge concept now allows us to consider another
type of invariance that previously seems to have been over looked.

A gauge transformation can often be associated with the evolutionary path of a
quantum mechanical object. This is not observable directly, a difference can
be such as when a quantum mechanical object can take distinct paths, and then
be recombined to demonstrate a phase difference such as in the Aharonov-Bohm
effect (A-B effect). This has been interpreted as a geometric phase in a
number of papers by Aharonov and various collaborators. Aharonov's career can
be viewed as largely devoted to elucidating what phase means in quantum
mechanics. The wavefunction is invariant under $U(1)$ gauge transformations,
so the wavefunction can be written with a constant relative phase
function,$\Lambda$, so we have:
\begin{equation}
\Psi\left(  x,t\right)  \rightarrow e^{i\Lambda}\Psi\left(  x,t\right)  .
\end{equation}
This can be interpreted as associating a memory with a path. In general, the
phase can depend upon the topological features of the phase such as associated
with the A-B effect\cite{Aharonov1959} . There is no reason, however, to limit
$\Lambda$, it can be an arbitrary function $\Lambda\left(  x\right)  $ or it
could also be considered a random variable with an underlying probability
density function (PDF).

\section{Random Gauge Transformation}

Under some circumstances this "randomization" of $\hat{\Lambda}$ could be used
to explicitly bring in randomness associated with a path with an underlying
PDF. Thus, it is possible to define a "random gauge" in quantum mechanics.
Random path-dependent effects become a stochastic memory associated with the
quantum mechanical system. We explore this is some detail and suggest an
experiment to test the effects of randomness on path dependent phase
difference effects.

Since phase has no physical meaning, there is no reason to associate it with a
fixed parameter $\theta_{\alpha}$, instead one could assume that phase to be a
random variable $\hat{\theta}_{\alpha}$ with an underlying PDF $P_{\theta
_{\alpha}}(\theta)$. Then one could claim that $\Psi$, when interpreted as
either a field coordinate or $\left\vert \Psi^{\alpha}\right\vert ,$ has the
usual Born probability interpretation in conventional quantum mechanics and
must be invariant under a random gauge transformation:
\begin{equation}
\hat{\Psi}\longmapsto e^{i\hat{\theta}_{\alpha}}\Psi.
\end{equation}
This is a new type of invariance to consider, for both the Schrodinger
equation or the equations of quantum field theory. This type of transform is
define to be a \textbf{type 1} \emph{random gauge transformation}. \ \ If one
associates randomness with gauge invariance rather than with $\Psi$ directly,
it opens some alternative ways to think about the interpretation of quantum
mechanics, which will be examined in a subsequent paper.

One is immediately led to the question of meaning for this \textbf{type 1}
random gauge transformation. \ One concept of random phase can be connected to
the idea of what a "path" could mean in quantum mechanics. Empirically a path
is associated with the propagation of disturbance which has an intensity
pattern that can localized to a measurement device located at a point $x$ and
a time $t$. It is given that propagation of a disturbance within that medium
depends on the positions of the point and the direction of propagation of the
disturbance. Based on the wave nature of both mechanical and optical waves, a
principle was proposed by Huygens' (and latter others) that explains how wave
fronts propagate in time\cite{Baker1987} . \emph{Huygens' Principle} underpins
the analysis of classical propagation problems in electromagnetic and
acoustics\cite{Beckmann1987} , and much of modern physics\cite{Feynman2010} .
Empirically, the propagation of disturbances at a given point in a medium
depends on the position of the point and the direction of propagation of the
disturbance. This means that\cite{Gelfand2000} :

\begin{enumerate}
\item \textit{Each point can be in one of two states: excited or at rest. No
concept of intensity of the distribution needs to be introduced.}

\item \textit{If a disturbance arrives at the point }$P$\textit{\ at the time
}$t_{2}$\textit{, then when starting at time }$t_{1}$\textit{\ }$(t_{2}\geq
t_{1})$\textit{, the point }$P$\textit{\ itself serves as a source of
disturbances propagating in the medium.}
\end{enumerate}

The mathematical form of Huygens' Principle can be expressed in a number of
different ways. We choose the following:

\begin{definition}
(Huygens' Principle): Given a wave front $\Psi(x,t)$, it is the propagator of
small waves which collectively make up the next wave front. The mathematical
form of this statement is:
\begin{equation}
\Psi(x,t_{2})=\int G(x,y)\,\Psi(y,t_{1})\,dy.\;\;\;\;\;\;(t_{2}>t_{1})
\end{equation}
$G(x,y)\doteq$ kernel of the propagator.
\end{definition}

The intensity pattern of the path at a latter time is the superposition of all
possible paths from $x$ to $y$. Now since there is no reason to choose a
particular path, we can associate it with the random gauge transformation
$\hat{\theta}_{\alpha}$ so we have $\hat{\theta}_{\alpha}=\hat{\theta}%
_{\alpha}(x)$. Thus, because there is a random gauge associated with $\Psi$,
there is an underling PDF associated with it as well. Our generalization is
based on some work in scattering theory and other applied physics
applications. Methods of determining the probability density function (PDF)
are well known for the transformation:
\begin{equation}
\hat{x}=a\sin\hat{\theta}%
\end{equation}
where $a$ is a deterministic parameter and $\hat{\theta}$ is random variable
with a uniform PDF\cite{Papoulis1991} $f_{\theta}(\hat{\theta})$. The PDF's
for this distribution and others are found in the Appendix. Given the
characterization of the probability density function of the angle it is
possible to determine the moments of the random variable:
\begin{equation}
\hat{y}_{c}=\sum_{i=1}^{N}\hat{r}_{i}\cos\hat{\theta}_{i}%
\end{equation}
or
\begin{equation}
\hat{y}_{s}=\sum_{i=1}^{N}\hat{r}_{i}\sin\hat{\theta}_{i}%
\end{equation}
for an arbitrary $\hat{r}_{i}$ and $\hat{\theta}_{i}$ provided their
characteristic functions exist\cite{Gray} . (Note these sums could also be
viewed as discrete random Fourier series\cite{Lanczos1975}\cite{Marcus1981} .)
Now, the two sums can be written as:
\begin{equation}
\hat{z}=\hat{y}_{c}+i\hat{y}_{s}=\sum_{j=1}^{N}\hat{A}_{j}e^{i\hat{\theta}%
_{j}}%
\end{equation}
which is the same mathematical form as the random gauge transformation
$\hat{\Psi}\longmapsto\sum_{\alpha}\hat{A}_{\alpha}e^{i\hat{\theta}_{\alpha}%
}\Psi$. Thus we have a means for associating different types of randomness
with the phase variable; some examples of such probability distributions are
the uniform distributions, Cauchy distributions, Normal distributions, etc. In
general, the mean $\left\langle \cdot\right\rangle $ of $\hat{z}$ can be
written as:
\begin{align}
\left\langle \hat{z}\right\rangle  & =\left\langle \hat{y}_{c}\right\rangle
+i\left\langle \hat{y}_{s}\right\rangle =\sum_{j=1}^{N}\left\langle \hat
{A}_{j}\right\rangle \left\langle e^{i\hat{\theta}_{j}}\right\rangle
\nonumber\\
& =\sum_{j=1}^{N}\left\langle \hat{A}_{j}\right\rangle \left\langle \cos
\hat{\theta}_{j}\right\rangle +i\sum_{j=1}^{N}\left\langle \hat{A}%
_{j}\right\rangle \left\langle \sin\hat{\theta}_{j}\right\rangle
\end{align}
provided the random variables in the sums are uncorrelated. (The appendix
considers specific PDF's for these angular variables.) Also, we have:%
\begin{align}
Var\left(  \hat{z}\right)   & =Var\left(  \hat{y}_{c}\right)  +iVar\left(
\hat{y}_{s}\right) \nonumber\\
& =\sum_{j=1}^{N}Var\left(  \hat{A}_{j}\right)  Var\left(  \cos\hat{\theta
}_{j}\right)  +i\sum_{j=1}^{N}Var\left(  \hat{A}_{j}\right)  Var\left(
\sin\hat{\theta}_{j}\right)  .
\end{align}
Thus, the mathematics of scattering theory can be drawn upon to explore how
the variances in the the paths between two paths can vary for different types
of random gauges. For example, most of the examples of Berry phase can be
readily generalized by exploring the variation in paths as instances of random
gauges. In the next section, we consider how to demonstrate this using a Berry
phase device. We then discuss how this might be related to variation in the
space-time metric as an instance of aspects of gravity as a gauge transformation.

It is possible to convert the $\sum_{i=1}^{N}r_{i}\cos\hat{\varsigma}_{i}$ to
a polynomial function $g(\theta,\varphi)$ using trigonometric identities:
\begin{equation}
\hat{G}(\hat{\theta},\hat{\varphi})=\sum_{j=1}^{N}r_{j}\cos\hat{\varsigma}%
_{j}.
\end{equation}
This formula can be interpreted as an antenna pattern, since any polynomial
function of variables, $\theta$ and $\varphi$ with $G\left(  \theta
,\varphi\right)  $ representing the gain of the antenna. In classical
electromagnetics, an antenna is the source to launch an electromagnetic wave
into space, and the antenna pattern characterizes how the electromagnetic wave
are distributed in the angular variables. This suggests a second type of
random gauge transformation. \ 

A \textbf{Type 2} \emph{random gauge transformation} associates the randomness
with the field coordinates so we have a family of PDF's associated with a
given gauge transformation:%
\begin{equation}
\hat{\Psi}\longmapsto\exp\left(  i\theta\left(  \alpha\left(  \hat{x}\right)
\right)  \right)  \Psi.
\end{equation}
It is in this form that Huygens' principle can be interpreted as a statement
in terms of antenna theory. Note this is done by interpreting $G(x,y)$ as
\ the antenna gain, which can be written as a function of angle variables:
\begin{equation}
G=G(\theta,\vartheta)=f(\sin\theta,\cos\vartheta)
\end{equation}
Now each point on a spherical surface of propagation can be treated as a point
source of outward propagating radiation, which, in the language of antenna
theory, is equivalent to the statement that the gain of the antenna is
$G(\theta,\varphi)=\frac{1}{2}$. If we assume that $G(\theta,\varphi)\neq
\frac{1}{2}$, then radiation expands non-uniformly. Thus we have non-radial
expansion of radiation from a point source. Huygens' principle with the gain
equal to one (inward and outward expansion) was more or less an observation
rather than an experimental fact or a theoretical principle etched in stone.
Thus, in line with the antenna theory interpretation one might argue that a
non-spherical expansion should be considered, so this would have the
functional form:
\begin{equation}
G(\theta,\vartheta)=\sum_{j=1}^{m}\sum_{i=1}^{n}a_{i}\cos\vartheta_{i}%
\sin\theta_{j}.
\end{equation}
The type 2 random gauge can be viewed as a generalization of the concept of an
antenna gain pattern. The field coordinates in quantum field theory can be
used as the source, in the same way the gain pattern of an antenna is, of
random fields that arise in space time.

\section{Physics Instances of Random Gauge Transformations}

The type 2 random gauge concept can be associated with the Feynman's path
integral and by using the random gauge concept in Huygens' principle. \ Given
a wave front $\Psi(x,t)$, which is the propagator of quantum waves, the gauge
acts collectively as a source to constitute the next wave front. The
mathematical form of this statement is:
\begin{equation}
\Psi(\vartheta,t_{2})=\int G(\hat{\theta},\hat{\vartheta})\,\Psi(\theta
,t_{1})\,d\theta.\;\;\;\;\;\;(t_{2}>t_{1})
\end{equation}
$G(\hat{\theta},\hat{\vartheta})\circeq$ random antenna pattern which
propagates $\Psi(\theta,t_{1})$ from the time $t_{1}$ to the time $t_{2}$. The
point of this is the randomness we normally associate with $\Psi$ in quantum
mechanics can instead be associated with mechanism for propagating $\Psi$ in
time if the most general form of gauge invariance is supposed for quantum
mechanics. It also provides a physical interpretation of the path integral by
a natural connection with Huygens' principle becoming the foundational
principle for quantum mechanics. Furthermore, if one takes this definition of
a propagator, then one can use it to explain the gauge invariance of $\Psi$,
and use this fact to explain phase invariance for many of the different
"paths" that can be taken in quantum mechanics as well as such things as the
Aharonov-Bohm effect and Berry phase.

The observation of Aharonov-Bohm phase\cite{Aharonov1959} suggests to one a
means of both interpreting and demonstrating some aspects of the random gauge
concept. In the Figure of the AB-effect, the electron wave function can be
written as\cite{Silverman2002} :
\begin{equation}
\Psi\left(  \mathbf{r},t\right)  =\Psi_{1}\left(  \mathbf{r},t\right)
e^{iS_{1}}+\Psi_{2}\left(  \mathbf{r},t\right)  e^{iS_{2}},
\end{equation}
where the numbers 1 and 2 refer to the slits and the phases for each path the
electron can take:%
\begin{equation}
S_{1,2}=\frac{e}{\hbar c}\int_{x_{0}}^{x_{1}}\mathbf{A}_{1.2}\cdot
d\mathbf{l}_{1,2}.
\end{equation}
The phase difference between the two components of the wave function is:%
\begin{equation}
S_{1}-S_{2}=\frac{e}{\hbar c}%
{\textstyle\oint\limits_{C}}
\mathbf{A}\cdot d\mathbf{l}=\frac{e\Phi}{\hbar c}%
\end{equation}
where $C$ is a closed contour about the solenoid and the magnetic flux $\Phi$
is the cumulative amount of flux in the interior of the solenoid. The solenoid
has a current passing through it. Instead of considering a normal current that
induces AB-phase difference, which assumes a current which is proportional to
$NI_{0}\cos\omega_{0}t$, where $N$ is the number of turns in the solenoid and
$\omega_{0}$ is the frequency of the current.%

\begin{figure}[ptb]%
\centering
\includegraphics[
natheight=3.634800in,
natwidth=4.802300in,
height=2.034in,
width=2.6783in
]%
{LI4YDE02.bmp}%
\end{figure}

If we consider the frequency no longer fixed, but disturbed by noise, it is
evident how noise can manifest itself in the AB-effect. If we specifically
modulate the current with noise, $\hat{I}=I_{0}\cos\left(  \omega_{0}t+\hat
{n}\right)  $, the effect of the noise is made manifest by an additional phase
difference in the phase difference between the two components of the wave
function. We have induced an effective random variation in the gauge path that
is equal to:%
\begin{equation}
\Delta\hat{S}=\hat{S}_{1}-\hat{S}_{2}=\frac{e\Phi\exp\left(  i\hat{\theta
}\right)  }{\hbar c}.
\end{equation}
The expected value of $\Delta\hat{S}$, $E\left[  \Delta\hat{S}\right]  $, is
proportional to $E\left[  \exp\left(  i\hat{\theta}\right)  \right]  $. For
many, but not all, probability distributions in the appendix, $E\left[
\exp\left(  i\hat{\theta}\right)  \right]  =0$. The probability distributions
for which $E\left[  \exp\left(  i\hat{\theta}\right)  \right]  \neq0$, are
those for which the characteristic function of the probability distribution of
$\theta$ is not even. The variance in the phase difference between the two
components of the wave function is:
\begin{equation}
Var\left[  \Delta\hat{S}\right]  =\frac{e\Phi}{\hbar c}\left[  Var\left(
\cos\hat{\theta}\right)  +iVar\left(  \sin\hat{\theta}\right)  \right]  .
\end{equation}
Thus, for zero mean Gaussian noise, the variance in the phase difference
between the two components of the wave function is $Var\left[  \Delta\hat
{S}\right]  =\frac{e\Phi}{\hbar c}\left[  \left(  1-e^{-2\sigma_{\theta}^{2}%
}\right)  \left(  1+i\right)  \right]  $ while for the Cauchy distribution, it
is $\frac{e\Phi}{\hbar c}\left[  \left[  1-e^{-2\alpha}\right]  \left(
1+i\right)  \right]  $. Higher order corrections can be found as well using
the material in the appendix. The same argument would also hold for both the
Aharonov-Casher effect and the gravitational equivalent to the AB-effect since
the phase difference between two paths for the quantum phenomena "$\circ$" can
be written as:
\begin{equation}
S_{1}-S_{2}=\frac{\alpha\Phi_{"\circ"}}{\hbar},
\end{equation}
where $\alpha$ is a phenomenological constant appropriate to the particular
situation and $\Phi_{"\circ"}$ is the flux associated with the phenomena
"$\circ$". An effective random variation in the gauge path for the phenomena
"$\circ$" is equal to:%
\begin{equation}
\Delta\hat{S}=\frac{\alpha\Phi_{"\circ"}e^{i\hat{\theta}_{"\circ"}}}{\hbar}.
\end{equation}

A third idea is couched in general relativity and related the space-time
metric. In general, the spatial metric of space-time can be expressed in terms
of direction cosines, for example in three dimensional space the
parameterization is:
\begin{align}
\hat{x}  & =r\cos\left(  \hat{\theta}\right)  \sin\left(  \hat{\varphi
}\right)  ,\\
\hat{y}  & =r\sin\left(  \hat{\theta}\right)  \sin\left(  \hat{\varphi
}\right)  ,\\
\hat{z}  & =r\cos\left(  \hat{\varphi}\right)  .
\end{align}
Now,
\[
s^{2}=x^{2}+y^{2}+z^{2}=\hat{x}^{2}+\hat{y}^{2}+\hat{z}^{2}=\hat{s}^{2}%
\]
so the distance is invariant with respect to a replace of the angles by random
variables, so:
\begin{equation}
s\rightarrow\hat{s}\rightarrow se^{i\sigma\left(  \theta,\varphi\right)  }.
\end{equation}
where $\sigma$ is an arbitrary function of $\theta$ and $\varphi$, so
$\sigma=\sigma\left(  \theta,\varphi\right)  $. Therefore, it is possible to
associate a fluctuation of the space-time metric random variations of in the
angular variables. These random variations are equivalent to a random gauge,
which can be observed using the idea of a path length phase difference for any
quantum phenomena. Any phenomena which has an invariant which it is possible
to replace $s\rightarrow\hat{s}$ as in Eq (27) potentially has a hidden random
gauge associated with it, so the formula for a gauge path for the phenomena
"$\circ$" can be used to find it:%
\begin{equation}
\Delta\hat{S}=\frac{\alpha\Phi_{"\circ"}e^{i\hat{\theta}_{"\circ"}}}{\hbar}.
\end{equation}
A Mach-Zenhender interferometer, combined with weak amplification device such
as has been discussed by Aharonov\cite{Aharonov2005} , might be used to detect
such random variations in the metric. In a latter publication, the random
gauge concept will be used to reexamine what Aharonov has termed modular momentum.

\section{Discussion and Conclusions}

Two concepts of random gauge invariance have been introduced in this paper
with separate examples of what random gauge could mean in a physical setting.
Random phase probability density functions are explained in the appendix with
examples of a variety of probability density functions. One concept of random
phase is associated with the Feynman's path integrals and used to provide
another interpretation of the Feynman path integral. Another interpretation of
random gauge is introduced using the Aharonov-Bohm effect and a method is
proposed for detecting the randomness is proposed. Additionally, a gauge
proposed which provides an explanation for fluctuations in the space-time
metric and a method is proposed for detecting it.

\textbf{Acknowledgement:} Thanks to Joshua Bellamy for reading this document
and comments that significantly improved the document.

\appendix

\section{Appendix: Superposition of Random Sinusoidal Functions}

The problem of random flights, dates back to a paper by Lord Rayleigh. (Note
these sums could be viewed as random Fourier
series\cite{Lanczos1975,Marcus1981} .) Our generalization is based on some
work in scattering theory and other applied physics applications. Methods of
determining the probability density function (PDF) are well known for the
transformation:
\begin{equation}
\hat{x}=a\sin\hat{\theta}%
\end{equation}
where $a$ is a deterministic parameter and $\hat{\theta}$ is a random variable
with a uniform PDF\cite{Papoulis1991} $f_{\theta}(\hat{\theta})$. Given the
characterization of the probability density function of the angle it is
possible to determine the moments of the random variable

The transformation $\hat{z}=A\sin(\hat{\varphi})$ a PDF $f_{\varphi}(\varphi)$
is onto but not one-to-one over the interval beyond $[-A,\,A]$. Thus it has an
infinite number of zeros. It is more convenient to determine the
characteristic function (CF) directly, so the Fourier transformation of the
PDF is:
\begin{equation}
M_{z}(\omega)=\left\langle e^{i\omega A\sin(\varphi)},f_{\varphi}%
(\varphi)\right\rangle .
\end{equation}
The exponential can be written as:
\begin{equation}
\sum_{n=-\infty}^{\infty}J_{n}(\omega A)e^{in\varphi}=e^{i\omega A\sin
(\varphi)},
\end{equation}
so the CF is given by:
\begin{equation}
M_{z}(\omega)=\sum_{n=-\infty}^{\infty}J_{n}(\omega A)F(n);
\end{equation}
where $F(n)$ is the Fourier transform of the PDF for the angle variable
$f_{\varphi}(\varphi)$ which is evaluated for $n$. Noting the Bessel functions
can be rewritten as ($J_{-n}(x)=(-)^{n}J_{n}(x)$).

\begin{proposition}
The CF of the transformation $\hat{z}=A\sin(\hat{\varphi})$ is:
\begin{equation}
M_{z}^{\sin}(\omega)=J_{0}(\omega A)F_{s}(0)+\sum_{n=1}^{\infty}J_{n}(\omega
A)S(n)
\end{equation}
where $S(n)=\left[  F(n)+(-)^{n}F(-n)\right]  $. \ 
\end{proposition}

\begin{proposition}
The CF for the$\ $transformation $\widehat{x}=A\cos(\hat{\varphi})=A\sin
(\hat{\varphi}-\frac{\pi}{2}),$ which amounts to replacing $\varphi$ by
$\varphi-\frac{\pi}{2}$ in the exponential, so the CF is:
\begin{equation}
M_{z}^{\cos}(\omega)=J_{0}(\omega A)F_{c}(0)+\sum_{n=1}^{\infty}J_{n}(\omega
A)C(n).
\end{equation}
where $C(n)=\left[  (-)^{n}F(n)+F(-n)\right]  $.
\end{proposition}

Expressions for the moments of the characteristic function can be found from
this formula:
\begin{equation}
\left\langle x^{m}\right\rangle =\frac{1}{i^{m}}\left.  \frac{\partial
^{m}M_{P}(\omega)}{\partial\omega^{m}}\right\vert _{\omega=0}=\frac{R^{m}%
}{i^{m}}2^{m}\{S(p)/C(p)\}\left.  J_{p}^{(m)}(\omega R)\right\vert _{\omega
=0}.
\end{equation}
This expression allows us to determine $\left\langle x^{n}\right\rangle $,
which requires us to know the n-th derivative of the Bessel function. Since
$J_{0}(0)=1$ and $J_{n}(0)=0$ for $n\neq0$,\ the only terms that remain after
we take the derivative with respect to $\omega$ and set it equal to zero are
those Bessel functions that have zero coefficients, e.g. those of the form
$J_{p-m}(x)$ which are one when $p=m$. This allows us to determine the moments
to arbitrary order using the recursion relation $J_{p-1}(x)-J_{p+1}%
(x)=2J_{p}^{\prime}(x)$ (' denotes derivative). We can continue with this
process to evaluate arbitrary derivatives of the Bessel functions to arbitrary
order as:
\begin{align}
2^{N}J_{p}^{(N)}(x)=J_{p-N}(x)+\left(  -\right)  ^{1}\binom{N}{1}J_{p+2-N}(x)
& +\binom{N}{2}J_{p+4-N}(x)+\left(  -\right)  ^{3}\binom{N}{3}J_{p+6-N}%
(x)+...+\;\nonumber\\
& \left(  -\right)  ^{N-1}\binom{N}{N-1}J_{p+N-2}(x)+\left(  -\right)
^{N}J_{p+N}(x),
\end{align}
where $J_{p}^{(N)}(x)$ denotes the $N^{th}$ derivative of the Bessel function
and $\binom{N}{m}$ is the binomial symbol where it is understood that
$\binom{N}{m}=0$ if $m>N$. Noting that $J_{0}(0)=1$ and $J_{n}(0)=0$ for $n$
not equal to zero, then only the even derivatives of $J_{0}(x)$ are not equal
to zero. The second moment is for both functions are:
\begin{equation}
\left\langle x^{2}\right\rangle =-\left.  \frac{\partial^{2}M_{\varphi}%
(\omega)}{\partial\omega^{2}}\right\vert _{\omega=0}=\frac{R^{2}}{4}\left[
2F(0)-\left(  F(2)+F(-2)\right)  \right]  .
\end{equation}

\begin{corollary}
\textbf{Gaussian or Normal Distribution:} The PDF of the zero mean normal
distribution has a CF given by:
\begin{equation}%
\begin{array}
[c]{c}%
\frac{1}{\sqrt{2\sigma_{\theta}^{2}\pi}}e^{-\frac{\theta^{2}}{2\sigma_{\theta
}2}}\Leftrightarrow e^{-\sigma_{\theta}^{2}\omega^{2}/2}\\
\text{thus}\\
\text{ }F(n)=e^{-\sigma_{\theta}^{2}n^{2}/2}%
\end{array}
\end{equation}
which is even. Note the mean is zero and the second moment for either
transformation is:
\begin{equation}
\left\langle x^{2}\right\rangle =\left.  \frac{\partial^{2}M_{\varphi}%
(\omega)}{\partial\omega^{2}}\right\vert _{\omega=0}=\frac{R^{2}}{2}\left[
1-e^{-2\sigma_{\theta}^{2}}\right]  .
\end{equation}
The third moment is zero, and the fourth moment for either transformation is:
\[
\left\langle x^{4}\right\rangle =\left.  \frac{\partial^{4}M_{\varphi}%
(\omega)}{\partial\omega^{4}}\right\vert _{\omega=0}=\frac{R^{4}}{8}\left[
e^{-8\sigma_{\theta}^{2}}-4e^{-2\sigma_{\theta}^{2}}+3\right]
\]
Thus, the CF's are:%
\begin{equation}
M_{\varphi}^{\sin}(\omega)=M_{\varphi}^{\cos}(\omega)=J_{0}(\omega
A)+2\sum_{m=1}^{\infty}J_{2m}(\omega A)e^{-2\sigma_{\theta}^{2}m^{2}}.
\end{equation}

\end{corollary}

\begin{corollary}
\textbf{Non-zero mean Gaussian:} The PDF of the nonzero mean normal
distribution has a CF given by:%
\begin{equation}
\sqrt{\frac{2}{\pi^{2}\sigma_{\theta}^{2}}}e^{-\omega^{2}\sigma_{\theta}%
^{2}/2}e^{-i\omega\theta_{0}}%
\end{equation}
thus:
\begin{equation}
F(n)=e^{-n^{2}\sigma_{\theta}^{2}/2}e^{-in\theta_{0}}.
\end{equation}
The mean is:%
\begin{equation}
\left\langle x\right\rangle =\left.  \frac{\partial M_{\varphi}(\omega
)}{\partial\omega}\right\vert _{\omega=0}=2e^{-\sigma_{\theta}^{2}/2}\left[
\cos\theta_{0}\right]  R,
\end{equation}
and the second moment is:
\begin{equation}
\left\langle x^{2}\right\rangle =\left.  \frac{\partial^{2}M_{\varphi}%
(\omega)}{\partial\omega^{2}}\right\vert _{\omega=0}=2\left[  1-e^{-2\sigma
_{\theta}^{2}}\cos\left(  2\theta_{0}\right)  \right]  R^{2},
\end{equation}
while the fourth moments is:
\begin{equation}
\left\langle x^{4}\right\rangle =\frac{R^{4}}{8}\left[  e^{-8\sigma_{\theta
}^{2}}\cos(4\theta_{0})-4e^{-2\sigma_{\theta}^{2}}\cos(2\theta_{0})+6\right]
.
\end{equation}

\end{corollary}

\begin{corollary}
\textbf{Laplace Distribution:} The PDF's CF is:
\begin{equation}
\frac{\alpha}{2}e^{-\alpha\left\vert x\right\vert }\Leftrightarrow\frac
{\alpha^{2}}{\alpha^{2}+\omega^{2}}\text{,}\;
\end{equation}
thus:
\begin{equation}
F(n)=\frac{\alpha^{2}}{\alpha^{2}+n^{2}},
\end{equation}
which is even. Thus the mean is zero and the second moment is:
\begin{equation}
\left\langle x^{2}\right\rangle =-\left.  \frac{\partial^{2}M_{\varphi}%
(\omega)}{\partial\omega^{2}}\right\vert _{\omega=0}=2R^{2}\left[
1-\frac{\alpha^{2}}{\alpha^{2}+4}\right]  .
\end{equation}
while the fourth moment is:
\begin{equation}
\left\langle x^{4}\right\rangle =\left.  \frac{\partial^{4}M_{\varphi}%
(\omega)}{\partial\omega^{4}}\right\vert _{\omega=0}=\frac{R^{4}}{8}\left[
\frac{\alpha^{2}}{\alpha^{2}+16}-\frac{4\alpha^{2}}{\alpha^{2}+4}+6\right]  .
\end{equation}

\end{corollary}

\begin{corollary}
\textbf{Cauchy\ Distribution:} The PDF's CF is:%
\begin{equation}
\frac{\alpha/\pi}{\alpha^{2}+x^{2}}\Leftrightarrow e^{-\alpha\left\vert
\omega\right\vert },
\end{equation}
thus%
\begin{equation}
F(n)=e^{-\alpha\left\vert n\right\vert },
\end{equation}
which is even. Thus we have the somewhat amusing result that the transformed
Cauchy has finite moments, while the original distribution doesn't. Note the
mean is zero and the second moment is:
\begin{equation}
\left\langle x^{2}\right\rangle =-\left.  \frac{\partial^{2}M_{\varphi}%
(\omega)}{\partial\omega^{2}}\right\vert _{\omega=0}=2A^{2}\left[
1-e^{-2\alpha}\right]  .
\end{equation}
while the fourth moment is:
\begin{equation}
\left\langle x^{4}\right\rangle =\left.  \frac{\partial^{4}M_{\varphi}%
(\omega)}{\partial\omega^{4}}\right\vert _{\omega=0}=\frac{R^{4}}{8}\left[
e^{-4\alpha}-4e^{-2\alpha}+6\right]  .
\end{equation}
\noindent Thus we have the somewhat amusing result that the transformed Cauchy
has finite moments, while the original distribution does not.
\end{corollary}

For products of random variables, it is still useful to know the PDF of the
one dimensional sinusoidal transform. Now if we apply the
identity\cite{Abramowitz1965} :
\begin{equation}
\int_{-\infty}^{\infty}e^{-i\omega x}J_{n}(x)\;dx=\frac{2(-i)^{n}T_{n}%
(\omega)}{\pi\sqrt{(1-\omega^{2})}}\Theta(1-\left\vert \omega\right\vert ).
\end{equation}
Note $T_{n}(x)$ is the n-th order Chebyshev polynomial which is defined as:
\begin{equation}
T_{n}(x)=\frac{n}{2}\sum_{m=0}^{\left[  n/2\right]  }(-)^{m}\frac{\left(
n-m-1\right)  !}{m!(n-2m)!}(2x)^{n-2m},
\end{equation}
$[\cdot]$ means the largest integer contained therein. Note all the subsequent
properties of Chebyshev polynomials are drawn from Arfken\cite{Arfken1985} .

\begin{proposition}
The PDF of the transformation $\hat{y}=A\sin(\hat{\theta})$ is:
\begin{equation}
f_{y}^{\sin\theta}(y)=\frac{2}{\pi}\frac{\left[  F(0)+\sum_{n=1}^{\infty
}U_{\sin(\theta)}(n)T_{n}(\frac{y}{A})\right]  }{\sqrt{(1-\left(  \frac{y}%
{A}\right)  ^{2})}},
\end{equation}
where:
\begin{equation}
U_{\sin(\theta)}(n)=\left[  F(n)+(-)^{n}F(-n)\right]  (-i)^{n}.
\end{equation}
This expression is only valid for $y\in\left[  -A,A\right]  $ and zero
elsewhere. Equivalently, we could multiply by $\Theta(1-\left\vert \frac{y}%
{A}\right\vert )$.
\end{proposition}

\begin{proposition}
The PDF of the coordinate transformation $\hat{y}=A\cos(\hat{\theta})$ is:
\begin{equation}
f_{y}^{\cos\theta}(y)=\frac{2}{\pi}\frac{\left[  F(0)+\sum_{n=1}^{\infty
}U_{\cos(\theta)}(n)T_{n}(\frac{y}{A})\right]  }{\sqrt{(1-\left(  \frac{y}%
{A}\right)  ^{2})}}%
\end{equation}
where:%
\begin{equation}
U_{\cos(\theta)}(n)=\left[  (-)^{n}F(n)+F(-n)\right]  .
\end{equation}
This expression is only valid for $y\in\left[  -A,A\right]  $ and zero elsewhere.
\end{proposition}

\begin{corollary}
\textbf{Triangular Distribution}: \ Since $F(n)=\frac{4\sin^{2}(na/2),}%
{a^{2}n^{2}}$, which is an even function, the PDF's for the sin transform is
given by:%
\begin{equation}
f_{y}^{\sin}(y)=\frac{2}{\pi}\frac{\left[  1+2\sum_{m=1}^{\infty}(-)^{m}%
\frac{\sin^{2}(ma)}{a^{2}m^{2}}T_{2m}(\frac{y}{A})\right]  }{\sqrt{\left(
1-\left(  \frac{y}{A}\right)  ^{2}\right)  }}\Theta(1-\left\vert \frac{y}%
{A}\right\vert ),
\end{equation}
while that for the cosine is:
\begin{equation}
f_{y}^{\cos}(y)=\frac{2}{\pi}\frac{\left[  1+2\sum_{m=1}^{\infty}\frac
{\sin^{2}(ma)}{a^{2}m^{2}}T_{2m}(\frac{y}{A})\right]  }{\sqrt{\left(
1-\left(  \frac{y}{A}\right)  ^{2}\right)  }}\Theta(1-\left\vert \frac{y}%
{A}\right\vert ).
\end{equation}

\end{corollary}

\begin{corollary}
\textbf{Gaussian or Normal Distribution}: Since $F(n)=e^{-\frac{n^{2}}%
{4\alpha}\text{ }}$, which is an even function, the PDF for the sin
transformation is:%
\begin{equation}
f_{y}^{\sin}(y)=\frac{2}{\pi}\frac{\left[  1+2\sum_{m=1}^{\infty}%
(-)^{m}e^{-\frac{m^{2}}{\alpha}}T_{2m}(\frac{y}{A})\right]  }{\sqrt{\left(
1-\left(  \frac{y}{A}\right)  ^{2}\right)  }}\Theta(1-\left\vert \frac{y}%
{A}\right\vert ).
\end{equation}
The PDF for the cosine transformation is:%
\begin{equation}
f_{y}^{\cos}(y)=\frac{2}{\pi}\frac{\left[  1+2\sum_{m=1}^{\infty}%
e^{-\frac{m^{2}}{\alpha}}T_{2m}(\frac{y}{A})\right]  }{\sqrt{\left(  1-\left(
\frac{y}{A}\right)  ^{2}\right)  }}\Theta(1-\left\vert \frac{y}{A}\right\vert
).
\end{equation}
This is a Gaussian sum, so a closed form evaluation of the sum is unknown by
current techniques.
\end{corollary}

\begin{corollary}
\textbf{Laplace Distribution}: Since $F(n)=\frac{\alpha^{2}}{\alpha^{2}+n^{2}%
}$, which is an even function, the PDF\ for the sin transformation is:%
\begin{equation}
f_{y}^{\sin}(y)=\frac{2}{\pi\sqrt{\left(  1-\left(  \frac{y}{A}\right)
^{2}\right)  }}\left[  1+2\sum_{m=1}^{\infty}(-)^{m}\frac{\alpha^{2}%
T_{2m}(\frac{y}{A})}{\alpha^{2}+4m^{2}}\right]  \Theta(1-\left\vert \frac
{y}{A}\right\vert ).
\end{equation}
The PDF for the cosine transformation is:%
\begin{equation}
f_{y}^{\cos}(y)=\frac{2}{\pi\sqrt{\left(  1-\left(  \frac{y}{A}\right)
^{2}\right)  }}\left[  1+4\sum_{m=1}^{\infty}\frac{\alpha^{2}T_{2m}(\frac
{y}{A})}{\alpha^{2}+4m^{2}}\right]  \Theta(1-\left\vert \frac{y}{A}\right\vert
).
\end{equation}

\end{corollary}

\begin{corollary}
\textbf{Cauchy\ Distribution}: \noindent Since $F(n)=e^{-\alpha\left\vert
n\right\vert }$, which is an even function, the PDF\ for the sin
transformation is:%
\begin{equation}
f_{y}^{\sin}(y)=\frac{2}{\pi}\frac{\left[  1+2\sum_{m=1}^{\infty}%
(-)^{m}e^{-2\alpha\left\vert m\right\vert }T_{2m}(\frac{y}{A})\right]  }%
{\sqrt{\left(  1-\left(  \frac{y}{A}\right)  ^{2}\right)  }}\Theta
(1-\left\vert \frac{y}{A}\right\vert ).
\end{equation}
The PDF for the cosine transformation is:%
\begin{equation}
f_{y}^{\cos}(y)=\frac{2}{\pi}\frac{\left[  1+4\sum_{m=1}^{\infty}%
e^{-2\alpha\left\vert m\right\vert }T_{2m}(\frac{y}{A})\right]  }%
{\sqrt{\left(  1-\left(  \frac{y}{A}\right)  ^{2}\right)  }}\Theta
(1-\left\vert \frac{y}{A}\right\vert ).
\end{equation}

\end{corollary}

Further results can be obtained without considering specific PDF's. The
orthogonality relationship for the Chebyshev polynomials\cite{Arfken1985} is:
\begin{equation}
\int_{-1}^{1}\frac{T_{m}(y)T_{n}(y)}{\sqrt{1-y^{2}}}dy=\frac{\pi}{2}%
\delta_{m,n}(n>0),
\end{equation}
or $\pi$ for $m=n=0$. Any integral of a polynomial function $f(x)$ with a
Chebyshev polynomial can be evaluated using:
\begin{equation}
x^{n}=\frac{1}{2^{n-1}}[T_{n}(x)+%
\genfrac{(}{)}{}{}{n}{1}%
T_{n-2}(x)+%
\genfrac{(}{)}{}{}{n}{2}%
T_{n-4}(x)+...],
\end{equation}
where the series terminates with $\binom{n}{m}T_{1}(x)$ for $n=2m+1$ or
$\frac{1}{2}\binom{n}{m}T_{0}(x)$ for $n=2m$. With these two results, the
means and standard deviations can be computed without specific knowledge of
the PDF's, since a specific $F(n)$ does not effect these calculations. The
mean is:
\begin{equation}
\overline{x}=\frac{AU_{\sin(\theta)/\cos(\theta)}(1)}{2}.
\end{equation}
The second moment is determined to be:
\begin{equation}
\overline{x}^{2}=\frac{A^{2}}{2}\left[  1+\frac{U_{\sin(\theta)/\cos(\theta
)}(2)}{2}\right]
\end{equation}
The standard deviation is therefore:
\begin{equation}
\sigma_{x}=\frac{A}{\sqrt{2}}\sqrt{1+\frac{U_{\sin(\theta)/\cos(\theta)}%
(2)}{2}-\frac{U_{\sin(\theta)/\cos(\theta)}^{2}(1)}{2}}.
\end{equation}
The $m^{th}$-moment is given by:
\begin{equation}
\overline{x}^{m}=\frac{A^{2l}}{2^{2l}}[U_{\sin(\theta)/\cos(\theta)}%
(2l)\binom{2l}{1}U_{\sin(\theta)/\cos(\theta)}(2\left(  l-1\right)
)+...+\binom{2l}{l}F(0)],
\end{equation}
for $m$ even, and the m-th moment is:
\begin{align}
\overline{x}^{m}  & =\frac{A^{2l+1}}{2^{2l+1}}[U_{\sin(\theta)/\cos(\theta
)}(2l+1)+\binom{2l+1}{1}U_{\sin(\theta)/\cos(\theta)}(2l-1)+...\nonumber\\
& +\binom{2l+1}{l}U_{\sin(\theta)/\cos(\theta)}(1)]
\end{align}
for $m$ odd.

\bibliographystyle{spiebib}
\bibliography{report}

\end{document}